\documentclass[showpacs,showkeys,11pt,
preprint,preprintnumbers,nofootinbib,
groupedaddress,superscriptaddress,amsmath,amssymb]{revtex4}

\usepackage[dvipdfm]{graphicx}
\usepackage[hypertex]{hyperref}

\usepackage{cancel,graphics,subfigure}
\usepackage{color}

\begin{document}
\title{
QCD Corrections to Neutron Electric Dipole Moment \\
from Dimension-six Four-Quark Operators
}
\preprint{IPMU12-0088}
\author{Junji Hisano}
\email{hisano@eken.phys.nagoya-u.ac.jp}
\affiliation{Department of Physics, Nagoya University, Nagoya 464-8602, Japan}
\affiliation{IPMU, TODIAS, University of Tokyo, Kashiwa 277-8568, Japan}
\author{Koji Tsumura}
\email{ko2@eken.phys.nagoya-u.ac.jp}
\affiliation{Department of Physics, Nagoya University, Nagoya 464-8602, Japan}
\affiliation{Department of Physics, National Taiwan University, Taipei 10617, Taiwan} 
\author{Masaki J.S. Yang}
\email{yang@eken.phys.nagoya-u.ac.jp}
\affiliation{Department of Physics, Nagoya University, Nagoya 464-8602, Japan}
\affiliation{Department of Physics, University of Tokyo, Tokyo 113-0033, Japan}

\date{\today}
\begin{abstract}
  In this Letter, the renormalization-group equations for the
  (flavor-conserving) CP-violating interaction are derived up to the
  dimension six, including all the four-quark
  operators, at one-loop level. We apply them to the models with the
  neutral scalar boson or the color-octet scalar boson which have CP-violating Yukawa
  interactions with quarks, and discuss the neutron electric dipole moment in
  these models.

\end{abstract}
\maketitle

\section{Introduction} 

The electric dipole moment (EDM) for neutrons is sensitive to CP
violation in physics beyond the standard model (SM) around TeV
scale. This is because, while the CP phase in the
Cabibbo-Kobayashi-Maskawa (CKM) matrix is $O(1)$, the CKM contribution
to the neutron EDM is too much suppressed \cite{Shabalin:1978rs} to be
observed in near future. (The recent evaluation of the CKM contribution
to the neutron EDM is given in Refs.~\cite{Mannel:2012qk}.)  The
naturalness problem in the Higgs-boson mass term in the SM might
require new physics at TeV scale, and many extensions of the SM
generically have CP-violating interactions.  The supersymmetric
standard model, which is the leading candidate for the TeV-scale
physics, is severely constrained from the EDM measurements
\cite{Pospelov:2005pr} .

The (flavor-conserving) CP-violating effective operators at parton
level up to the dimension six are the QCD theta term, the EDMs and the
chromoelectric dipole moments (CEDMs) of quarks, the Weinberg's
three-gluon operator~\cite{Ref:GGG-Op} and the four-quark
operators. In the evaluation of the neutron EDM, the CP-violating
four-quark operators tend to be ignored since the four-light-quark
operators suffer from chiral suppression in many models. However, the
four-quark operators including heavier ones, such as bottom/top
quarks, may give sizable contributions to the neutron EDM.  The EDMs,
CEDMs, and the three-gluon operator are radiatively generated from
the four-quark operators by integrating out heavy quarks.

In the multi-Higgs models, the Barr-Zee diagrams are known to give the
sizable contribution to the neutron EDM \cite{Barr:1990vd}. In the
Barr-Zee diagrams the heavy-quark loops are connected to light-quark
external lines by the neutral scalar boson exchange so that the CEDMs 
for light quarks are generated at two-loop level at $O(\alpha_s)$. However, 
it is not clear which renormalization scale should be chosen for
$\alpha_s$. In addition, the contributions from the Barr-Zee diagrams
at two-loop level to the quark EDMs vanish at $O(\alpha_s)$. However, 
it is still unclear that the higher-order corrections to the quark EDMs are 
negligible in the neutron EDM evaluation.

In this Letter, in order to answer those questions, we derive the
renormalization-group equations (RGEs) for the Wilson coefficients for
the CP-violating effective operators up to the dimension six at
one-loop level, including operator mixing.  The RGEs for the EDMs and
CEDMs for quarks and the three-gluon operator have been derived in
Ref.~\cite{Ref:SVZ,Ref:RGE-GGG,Ref:BGTW}. The next-leading order
corrections to them are also partially included \cite{Ref:DMFS}. We
include the four-quark operators in the calculation at the leading
order. Using the derived RGEs, we evaluate the EDMs and CEDMs for 
light quarks and the three-gluon operators induced by the
neutral scalar boson exchange including the QCD correction. We also discuss
the four-quark operators induced by the color-octet scalar boson.

This Letter is organized as follows. In the next section, we review the
neutron EDM evaluation from the parton-level effective Lagrangian at
the hadron scale. In Section 3, we derive RGEs for the Wilson
coefficients for the CP-violating effective operators up to the
dimension six at one-loop level. In Section 4, we show the effect of
the running $\alpha_s$ on the evaluation of the Wilson coefficients,
assuming the neutral scalar boson  exchange induces the CP-violating
effective operators.  In Section 5, another example of the application
of the RGEs is shown, assuming the effective operators induced by
a color-octet scalar boson.  Section 6 is devoted to conclusion.


\section{Neutron EDMs}
First, we review about evaluations of the neutron EDM from the low-energy
effective Lagrangian at parton level. The CP-violating interaction at
parton level around the hadron scale ($\mu_H^{}=1$~GeV) is given by
\begin{align}
 {\cal L}_{\rm CPV}
 =&\quad \theta \frac{\alpha_s}{8\pi}G^A_{\mu\nu}\widetilde{G}^{A\mu\nu}\nonumber
  \\
&-\frac{i}{2}\sum_{q=u,d,s}d_q\, \overline{q}(F\cdot\sigma)\gamma_5q
-\frac{i}{2}\sum_{q=u,d,s}\tilde{d}_q\, \overline{q}g_s(G\cdot\sigma)\gamma_5q
\nonumber \\
&+\frac{1}{3}w f_{ABC}G^{A}_{\mu\nu}\tilde{G}^{B\nu\lambda}
G^{C\mu}_\lambda.
\label{Lagrangian}
\end{align}
Here, $F_{\mu\nu}$ and $G^A_{\mu\nu} (A=1$--$8)$ are the electromagnetic and
gluon field strength tensors, $g_s$ is the strong coupling constant
($\alpha_s=g_s^2/4\pi)$, $F\cdot\sigma\equiv
F_{\mu\nu}\sigma^{\mu\nu}$, $G\cdot\sigma\equiv
G^A_{\mu\nu}\sigma^{\mu\nu}T^A$, and $\tilde{G}^A_{\mu\nu}\equiv
\frac{1}{2}\epsilon_{\mu\nu\rho\sigma}G^{A\rho\sigma}$ with
$\sigma^{\mu\nu}=\frac{i}2[\gamma^\mu,\gamma^\nu]$ and $\epsilon^{0123}=+1$. 
The matrix $T^A$ denotes the generators in the SU(3)$_C$ algebra, 
and $f^{ABC}$ is the structure constant. The first, second, third and forth terms in
Eq.~(\ref{Lagrangian}) are called the QCD $\theta$ term, the EDM and the CEDM 
for quarks, and the three-gluon operator, respectively. In this Letter, the covariant
derivative is defined as $D_{\mu}=\partial_{\mu}-ie Q_qA_\mu -ig_sG^A_\mu T^A$,
in which $A_\mu$ and $G^A_\mu$ are gauge fields for U(1)$_{EM}$ and
SU(3)$_{C}$, respectively with $Q_q$, the QED charge
($(Q_u,Q_d,Q_s)=(2/3,-1/3,-1/3))$. In Eq.~(\ref{Lagrangian}), we
ignore the CP-violating four-quark operators, since their coefficients
are often proportional to the light-quark masses in typical models, as
mentioned in the Introduction.

The neutron EDM is evaluated from the low-energy interaction at
parton level with the naive dimensional analysis, the chiral
perturbation theory, and the QCD sum rules, though they are considered
to have large uncertainties. The evaluation in term of the QCD sum rules is more
systematic than the others, at least for the contributions from the
QCD theta term, and the quark EDMs and CEDMs to the neutron EDM
\cite{hep-ph/9904483}. The recent evaluation of the neutron EDM with 
the QCD sum rules \cite{Hisano:2012sc} is 
\begin{align}
 d_n \simeq 2.9\times 10^{-17}\bar{\theta}~[e~{\rm cm}]+0.32d_d -0.08d_u +
e(+0.12\tilde{d}_d-0.12\tilde{d}_u-0.006\tilde{d}_s)~.
\label{sumrule}
\end{align}
In the evaluation, the recent QCD lattice result is used for the
low-energy constant $\lambda_n$, which is defined by $ \langle 0 |
\eta_n(x)|N(\vec{p},s)\rangle = \lambda_n u_n(\vec{p}, s)$ with
$\eta_n(x)$ the neutron-interpolating field.
If a value of $\lambda_n$ evaluated with the QCD sum rules is used,
the neutron EDM is enhanced by about five times compared with Eq.~(\ref{sumrule}). 

The contribution from the three-gluon operator might be comparable to
the quark EDMs and CEDMs. The quark EDMs and CEDMs are proportional
to the quark masses, while the three-gluon operator does not need to
suffer from chirality suppression. However, the size of the contribution
from the three-gluon operator depends on the methods of
the evaluation. In Ref.~\cite{Demir:2002gg} the authors compare the
several evaluations and propose
\begin{align}
d_n(w)\sim& ({\rm 10-30})\, {\rm MeV} \times  e w \,.
\label{weinbergopn}
\end{align}

\section{Operator Bases and Anomalous Dimension Matrix} 

We would like to introduce heavy quarks in the low-energy effective theory 
and evaluate their contributions to the neutron EDM. 
In this section, we show the one-loop RGEs for the
Wilson coefficients for the CP-violating effective operators up to the
dimension six, including heavy quarks.

First, we define the operator bases for the RGE analysis.  The
flavor-conserving effective operators for the CP violation
in QCD are given up to the dimension six as
\begin{align}
{\cal L}_\text{CPV} =& 
\sum_{i=1,2,4,5} \sum_q C_i^q(\mu) {\mathcal O}_i^q(\mu)
+C_3(\mu) {\mathcal O}_3(\mu) \nonumber \\
&+\sum_{i=1,2} \sum_{q'\ne q} \widetilde{C}_i^{q'q}(\mu) \widetilde{\mathcal O}_i^{q'q}(\mu)
+\frac12\sum_{i=3,4} \sum_{q'\ne q} \widetilde{C}_i^{q'q}(\mu) \widetilde{\mathcal O}_i^{q'q}(\mu)\,,
\label{effop}
\end{align}
where the sum of $q$ runs not only light quarks but also heavy ones,  and
we ignore the QCD theta term since it is irrelevant to our discussion here~\footnote{
The QCD theta term does not contribute to the RGEs for other CP-violating terms. Furthermore, there may be contribution to the QCD theta term from other CP violation terms, while the QCD theta term vanishes dynamically if the Peccei-Quinn symmetry is invoked. 
}.  
The effective operators are defined as 
\begin{align}
{\mathcal O}_1^q
=& -\frac{i}2 m_q \bar{q} \,eQ_q (F\cdot 
\sigma) \gamma_5q\, , \nonumber\\
{\mathcal O}_2^q
=& -\frac{i}2 m_q \overline{q} \,g_s (G\cdot\sigma) \gamma_5q \, , \nonumber\\
{\mathcal O}_3
=& -\frac16 g_s f^{ABC} \epsilon^{\mu\nu\rho\sigma} 
G^A_{\mu\lambda} {G^B}_{\nu}^{~\lambda} G^C_{\rho\sigma}\,,
\end{align}
and 
\begin{align}
{\mathcal O}_4^q
=&\, \overline{q_\alpha} q_\alpha \overline{q_\beta} \,i\gamma_5 q_\beta\,, \nonumber\\
{\mathcal O}_5^q
=&\, \overline{q_\alpha} \sigma^{\mu\nu} q_\alpha \overline{q_\beta} \,i\sigma_{\mu\nu}\gamma_5 q_\beta\,, \nonumber\\
\widetilde{\mathcal O}_1^{q'q}
=&\, \overline{q'_\alpha} q'_\alpha \overline{q_\beta} \,i\gamma_5 q_\beta\,, \nonumber\\
\widetilde{\mathcal O}_2^{q'q}
=&\, \overline{q'_\alpha} q'_\beta \overline{q_\beta} \,i\gamma_5 q_\alpha\,, \nonumber\\
\widetilde{\mathcal O}_3^{q'q}
=&\, \overline{q'_\alpha} \sigma^{\mu\nu} q'_\alpha \overline{q_\beta} \,i\sigma_{\mu\nu}\gamma_5 q_\beta\,, \nonumber\\
\widetilde{\mathcal O}_4^{q'q}
=&\, \overline{q'_\alpha} \sigma^{\mu\nu} q'_\beta \overline{q_\beta} \,i\sigma_{\mu\nu}\gamma_5 q_\alpha \,.
\label{4foperator}
\end{align}
Here, $m_q$ are masses for quark $q$. 
In Eq.~(\ref{4foperator}) we explicitly show the color indices, $\alpha$ and $\beta$. 
A factor of $1/2$ appears in front of the fourth term of Eq.~(\ref{effop}), 
since the term is symmetric under the exchange of $q'$ and $q$. 
The Wilson coefficients in Eq.~(\ref{effop}) are related to the parameters in Eq.~(\ref{Lagrangian}) as 
\begin{align}
d_q =&\, { m_{q}} \,e Q_q\, C^q_1(\mu_H^{})\, ,\nonumber\\
\tilde{d}_q =&\, {m_{q}}\, C^q_2(\mu_H^{})\, ,\nonumber\\
w=& -\frac12 g_s\, C_3(\mu_H^{})\, .
\end{align}

The RGEs for the Wilson coefficients of these operators are given as follows,
\begin{align}
\mu \frac{\partial}{\partial\mu}{\bf C}= {\bf C}{\bf \Gamma},
\end{align}
where the Wilson coefficients are written in a column vector as 
\begin{align}
{\bf C}=(C_1^q,C_2^q,C_3,C_4^q,C_5^q,
\widetilde{C}_1^{q'q},\widetilde{C}_2^{q'q},
\widetilde{C}_1^{qq'},\widetilde{C}_2^{qq'},
\widetilde{C}_3^{q'q},\widetilde{C}_4^{q'q}).
\end{align}
The anomalous dimension matrix is calculated at one-loop level as
\begin{align}
{\bf \Gamma} = \begin{bmatrix}
\frac{\alpha_s}{4\pi} \gamma_s & {\bf 0}                        & {\bf 0} \\
\frac1{(4\pi)^2} \gamma_{sf}   & \frac{\alpha_s}{4\pi} \gamma_f & {\bf 0} \\
\frac1{(4\pi)^2} \gamma'_{sf}  & {\bf 0}                        & \frac{\alpha_s}{4\pi} \gamma'_f
\end{bmatrix}, \label{Eq:Gamma}
\end{align}
where
\begin{align}
\gamma_s &=
\begin{bmatrix}
+8C_F & 0         & 0 \\
+8C_F & +16C_F-4N & 0 \\
0     & +2N       & N+2n_f+\beta_0 
\end{bmatrix}, \label{Eq:Gamma_s}
\end{align}
\begin{align}
\gamma_f &=
\begin{bmatrix}
-12C_F+6         & +\frac1N-\frac12 \\
+\frac{48}{N}+24 & +4C_F+6
\end{bmatrix}, \label{Eq:Gamma_f}
\end{align}
\begin{align}
\gamma'_f &=
\begin{bmatrix}
-12C_F        & 0                   & 0             & 0                   & +\frac1{N} & -1 \\
-6            & +\frac6{N}          & 0             & 0                   & -\frac12   & -C_F+\frac1{2N} \\
0             & 0                   & -12C_F        & 0                   & +\frac1{N} & -1 \\
0             & 0                   & -6            & +\frac6{N}          & -\frac12   & -C_F+\frac1{2N} \\
+\frac{24}{N} & -24                 & +\frac{24}{N} & -24                 & +4C_F      & 0 \\
-12           & -24C_F+\frac{12}{N} & -12           & -24C_F+\frac{12}{N} & +6         & -8C_F-\frac6{N}
\end{bmatrix}, \label{Eq:Gamma_f'}
\end{align}
\begin{align}
\gamma_{sf} &=
\begin{bmatrix}
+4      & +4  & 0 \\
-32N-16 & -16 & 0 
\end{bmatrix}, \label{Eq:Gamma_sf}
\end{align}
and
\begin{align}
\gamma'_{sf} &=
\begin{bmatrix}
0                                        & 0                     & 0 \\
0                                        & 0                     & 0 \\
0                                        & 0                     & 0 \\
0                                        & 0                     & 0 \\
-16N\frac{m_{q'}}{m_q}\frac{Q_{q'}}{Q_q} & 0                     & 0 \\
-16\frac{m_{q'}}{m_q}\frac{Q_{q'}}{Q_q}  & -16\frac{m_{q'}}{m_q} & 0
\end{bmatrix}, \label{Eq:Gamma_sf'}
\end{align}
where $C_F(=(N^2-1)/(2N))$ is the Casimir constant of the fundamental representation, 
$N(=3)$ is the number of the color, $n_f$ is the number of light flavor quarks, and 
$\beta_0(=11/3\times N-2/3\times n_f)$ is the leading-order beta function 
of strong coupling constant.

The anomalous dimensions for the dimension-five operators are
calculated in Ref.~\cite{Ref:SVZ}, and that for the three-gluon
operator is calculated by Ref.~\cite{Ref:RGE-GGG}. The mixings among
the dimension-five operators and the three-gluon operator are found in
Ref.~\cite{Ref:BGTW}. We newly calculate other terms in the anomalous
dimensions matrix, which are related with the four-quark
operators.  We note that the operator mixings between the EDM or CEDM
operators and the four-quark operators are generated at ${\mathcal
  O}(\alpha_s^0)$. 

\section{Neutral Scalar Boson Exchange}

In multi-Higgs models, a color-singlet neutral scalar boson $\phi$ may have
the CP-violating Yukawa coupling with quarks.  If the Yukawa
interaction violates the CP invariance, the CP-violating four-quark
operators are induced at tree level, after integrating the neutral scalar boson out, as
\begin{align}
C_4^q &= \sqrt2G_F 
\frac{m^2_q}{m_\phi^2}f_S^qf_P^q\,, \nonumber\\
\widetilde{C}_1^{q'q} &= \sqrt2G_F 
\frac{m_qm_{q'}}{m_\phi^2}f_S^qf_P^{q'}\,, \nonumber\\
\widetilde{C}_1^{qq'} &= \sqrt2G_F 
\frac{m_qm_{q'}}{m_\phi^2} f_S^{q'}f_P^q\,,
\end{align}
where we assume that $\phi$  is heavier than heavy quarks ($m_\phi\gg m_q,m_{q'}$).
Here, $f_S^q$ and $f_P^q$ are the CP-even and odd Yukawa coupling
constants, respectively,  defined as
\begin{align}
{\mathcal L}_\phi 
&= 2^{1/4}G_F^{1/2} m_q
\overline{q_\alpha} (f_S^q + i\,f_P^q \gamma_5) q_\alpha \phi,
\label{yukawa}
\end{align}
where $\phi$ is a (CP-even) real scalar field, and 
$G_F$ is the Fermi constant. We parametrize the
Yukawa coupling constants as they are proportional to the quark masses. 
In typical new physics models, these Yukawa coupling constants are 
taken to be proportional to masses of quarks in order to avoid the stringent 
experimental constraints from the flavor physics data. 
For the SM Higgs boson, the Yukawa 
coupling constants are of $f_S^q=1$ and $f_P^q=0$.
For the multi-Higgs models, {\it e.g.}, two-Higgs-doublet models, 
the coefficients $f_{S/P}^q$ may be much larger than unity because of 
the enhancement factor originated from the ratio of vacuum expectation values. 

It is known that, in these models, the EDMs and CEDMs for light
quarks are generated by the Barr-Zee diagrams at two-loop level, and
the three-gluon operator is also induced by the heavy-quark loops at
two-loop level. Let us derive those contributions using the RGEs for the
Wilson coefficients.

In the leading-logarithmic approximation, since $C_4^q$ is non-zero,
the coefficients for the EDM and the CEDM operators are generated from
the one-loop RGEs as
\begin{align}
C_1^q  =C_2^q
&= -\frac1{4\pi^2} C_4^q \ln\frac{m_\phi}{m_q} \,. 
\end{align}
These contributions are compared with the explicit calculation of 
one-loop diagrams with the scalar boson exchange as 
\begin{align}
C_1^q  =C_2^q
&= -\frac1{4\pi^2} C_4^q \left(\ln\frac{m_\phi}{m_q} -\frac34\right) \,,
\label{full1loop}
\end{align}
where a limit of $m_q \ll m_\phi$ is taken. 
The second term in the parentheses should be considered as the
short-distance contribution in which the loop momentum is around
$m_\phi$. 

The one-loop contributions from $C_4^q$ to the EDMs and CEDMs for
light quarks $(q=u,d,s)$ are negligible in the neutron EDM since they
are suppressed by powers of the light-quark masses.  However, when the
CEDMs for heavy quarks are generated, the
three-gluon operator is induced by integration of heavy quarks as follows \cite{Chang:1991ry},
\begin{align}
C_3(m_q)&=\frac{\alpha_s(m_q)}{8\pi} C^q_2(m_q) \,.
\label{3g}
\end{align}
Thus, from Eqs.~(\ref{full1loop}, \ref{3g}), we get
\begin{align}
C_3 &=
-\frac{\alpha_s}{32\pi^3} 
C_4^q \left(\ln\frac{m_\phi}{m_q} -\frac34\right)\,.
\end{align}
The result is consistent with the explicit calculation of the two-loop diagrams
for the three-gluon operator \cite{Dicus:1989va}. 

When $\widetilde{C}_1^{q'q}$ and/or $\widetilde{C}_1^{qq'}$ are
non-zero, the contribution to the CEDMs for light quarks
is derived at the two-loop level, using the RGEs for the 
the Wilson coefficients, Eq.~(\ref{Eq:Gamma}),  as
\begin{align}
C_2^q
&= \frac{\alpha_s}{8\pi^3} \frac{m_{q'}}{m_q} \Big(\ln\frac{m_\phi}{m_{q'}}\Bigr)^2
\Bigl[\widetilde{C}_1^{q'q}+\widetilde{C}_1^{qq'}\Bigr]\, .
\label{barr-zee}
\end{align}
This is because $\widetilde{C}_1^{q'q}$ and $\widetilde{C}_1^{qq'}$
are mixed with $\widetilde{C}_4^{q'q}$ and $\widetilde{C}_4^{qq'}$
which induce the CEDMs. When the Yukawa coupling constants in
Eq.~(\ref{yukawa}) are proportional to the quark masses, the induced
CEDMs for light quarks are not suppressed by their masses,
compared with the one-loop contribution.  The result in
Eq.~(\ref{barr-zee}) is consistent with the explicit calculation of the
Barr-Zee diagrams~\cite{Barr:1990vd} in a limit of $m_\phi \gg
m_q,m_{q'}$.
It is known that the contribution to the neutron EDM from the Barr-Zee diagram 
at  two-loop can be competitive with those from the one-loop diagrams~\cite{Barr:1990vd} . 
Therefore, the four-quark operator for heavy quarks can be important in some cases.

On the other hand, the EDMs for light quarks are not generated by
two-loop level diagrams of $O(\alpha_s)$, even if
$\widetilde{C}_1^{q'q}$ and $\widetilde{C}_1^{qq'}$ are non-zero. The
EDMs for light quarks have contributions from $\widetilde{C}_3^{q'q}$
($\widetilde{C}_3^{qq'}$) and $\widetilde{C}_4^{q'q}$
($\widetilde{C}_4^{qq'}$) in Eq.~(\ref{Eq:Gamma}). Their contributions
are exactly canceled with each others so that the EDMs vanish at the order.
This is also consistent with the explicit calculation of the Barr-Zee
diagrams.  The EDMs generated at two-loop level are suppressed by
$\alpha$.  However, the running effect of the strong coupling constant
may prevent the cancellation so that the EDM would be enhanced.

Now, let us consider the effect of the running strong coupling
constant, $\alpha_s(\mu)$.  Here we compare values of the EDM and CEDM operators 
for down quark and the three-gluon operators including and not
including the renormalization-group evolution of the strong coupling constant. We assume that
the Yukawa coupling constants for down and bottom quarks with
$\phi$ are non-zero in Eq.~(\ref{yukawa}) and then
\begin{align}
&\widetilde{C}_1^{bd}(m_\phi)\ne0, \quad C_4^b(m_\phi)\ne0\,, \nonumber  \\
&C_1^b(m_\phi)=C_2^b(m_\phi)= + \frac{3}{16\pi^2}C_4^b(m_\phi) \,.
\end{align}
The last assumption comes from Eq.~(\ref{full1loop}). 

In Fig.~1 the CEDM for down quark, $\tilde{d}_d$, (a) 
and the coefficient of the three-gluon operators, $w$, (b) 
at the hadron scale ($\mu=\mu_H^{}=1$~GeV) 
are shown as functions of $m_\phi^{}$ with
$f_S^d=f_P^d=1$ and $f_S^b=f_P^b=1$.  
Here, we ignore the contributions from top quark, and other short-distance effects.
If the scalar mass $m_\phi^{}$ is larger than the top quark mass ($m_\phi^{}> m_{t}$), 
the RGEs are solved using $\beta_{0}$ with $n_{f} = 6$, or if not, with $n_{f} = 5$.
When bottom quark is integrated out, the Wilson coefficient of Weinberg operator emerges.
Then the RGEs are solved using $\beta_{0}$ with $n_{f} = 4$ to the scale $\mu = m_{c}$, and 
with $n_{f} = 3$ to the scale $\mu = 1$ GeV. 
We use $m_d(\mu_H)=9$~MeV,
$m_c(m_c)=1.27$~GeV, $m_b(m_b)=4.25$~GeV, $m_t(m_t)=172.9$~GeV, and $\alpha_s(m_Z^{})=0.12$.
For the coefficient $w$, we multiply 10~MeV in the figure, which is a
factor in Eq.~(\ref{weinbergopn}), so that one may estimate the
contribution to the neutron EDM. It is from
Eqs.~(\ref{sumrule},\ref{weinbergopn}) found that the three-gluon
operator might be comparable to the CEDM when $f_{S/P}^d\sim f_{S/P}^b$.

In Fig.~2 the ratios of the CEDM for down quark (a) and the three-gluon
operator (b) at $\mu = m_b$ between including the running effect of
$\alpha_s$ and not including it (using the constant coupling
$\alpha_s=\alpha_s(m_b^{})$), are shown as functions of $m_\phi^{}$. It is found
that the  running coupling $\alpha_s(\mu)$ changes the CEDM
by about 20\% while the three-gluon operator is changed by at most 10
\%. These results come from inclusions of the four-quark operators to
the RGEs for the Wilson coefficients.

In Fig.~3 the ratio of the EDM and CEDM for down quark is presented as a
function of $m_\phi^{}$. 
The non-zero value of the EDM is generated as we mentioned.
It is also found that the ratio is roughly 
proportional to $\log m_\phi^{}$, and the absolute value is about 0.14. 
In the evaluation of the neutron EDM with the QCD sum rules, 
the size of the contribution from the down-quark EDM is $30$--$40$ \% 
of that from the CEDM. 

\begin{figure}[h]
\begin{center}
\begin{tabular}{ccc}
   \includegraphics[width=8cm,clip]{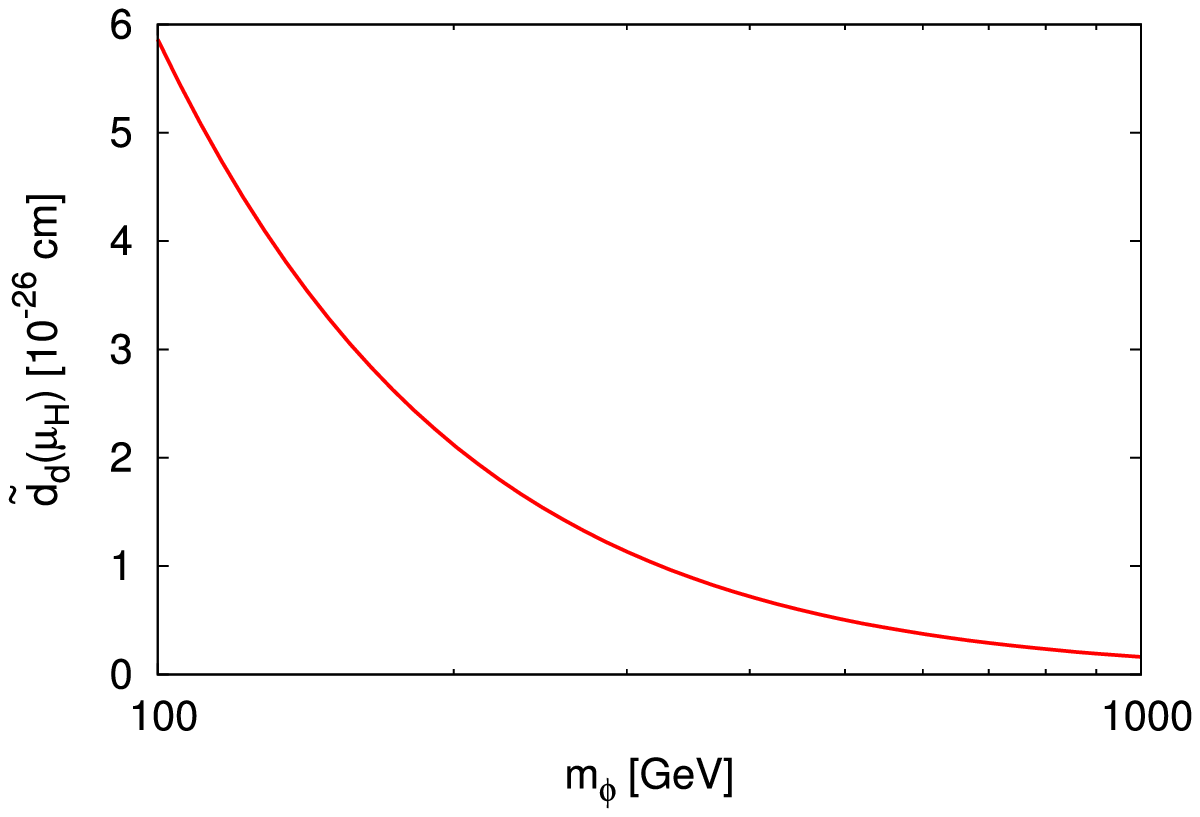} &
   \includegraphics[width=8cm,clip]{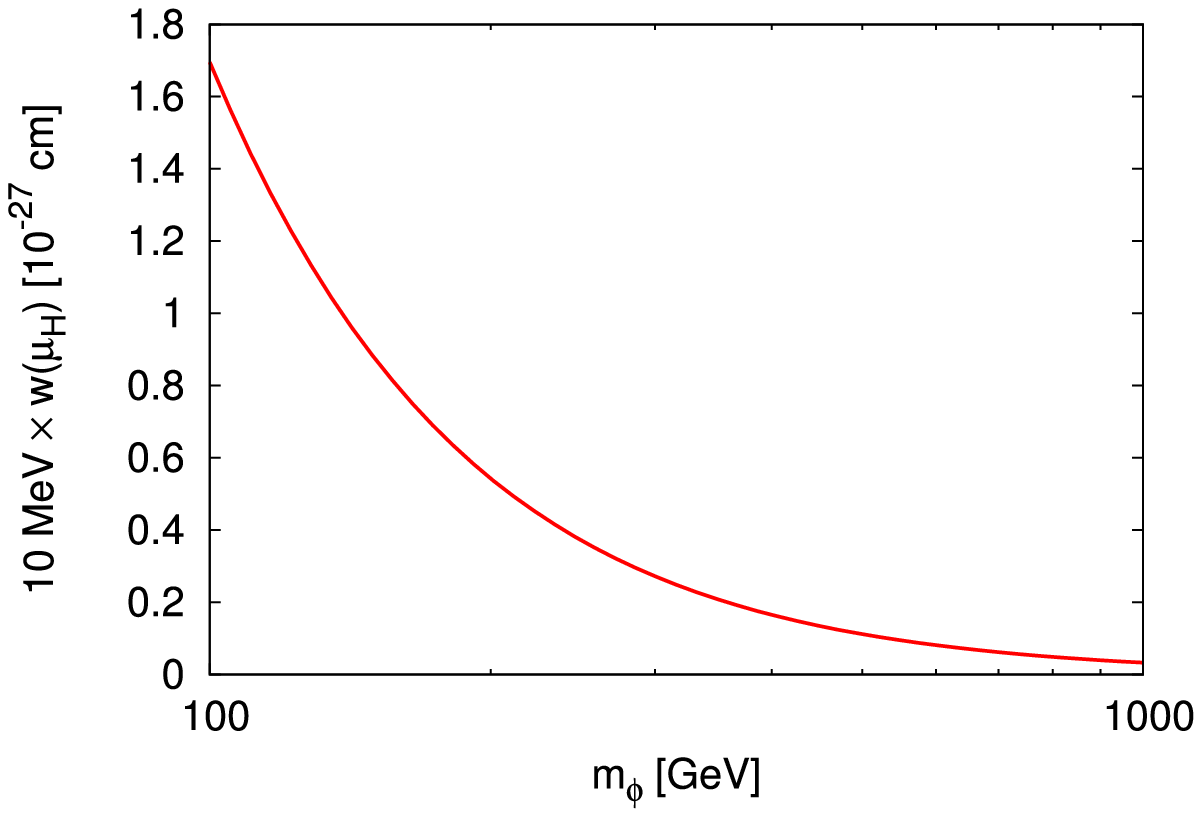} \\
   (a) & (b) 
\end{tabular} 
\caption{(a)  CEDM for down quark, $\tilde{d}_d$, 
  and  (b) coefficient of three-gluon operator, $w$, 
at hadron scale as functions of $m_\phi^{}$.}
\end{center}
\end{figure}

\begin{figure}[tb]
\begin{center}
\begin{tabular}{ccc}
   \includegraphics[width=8cm,clip]{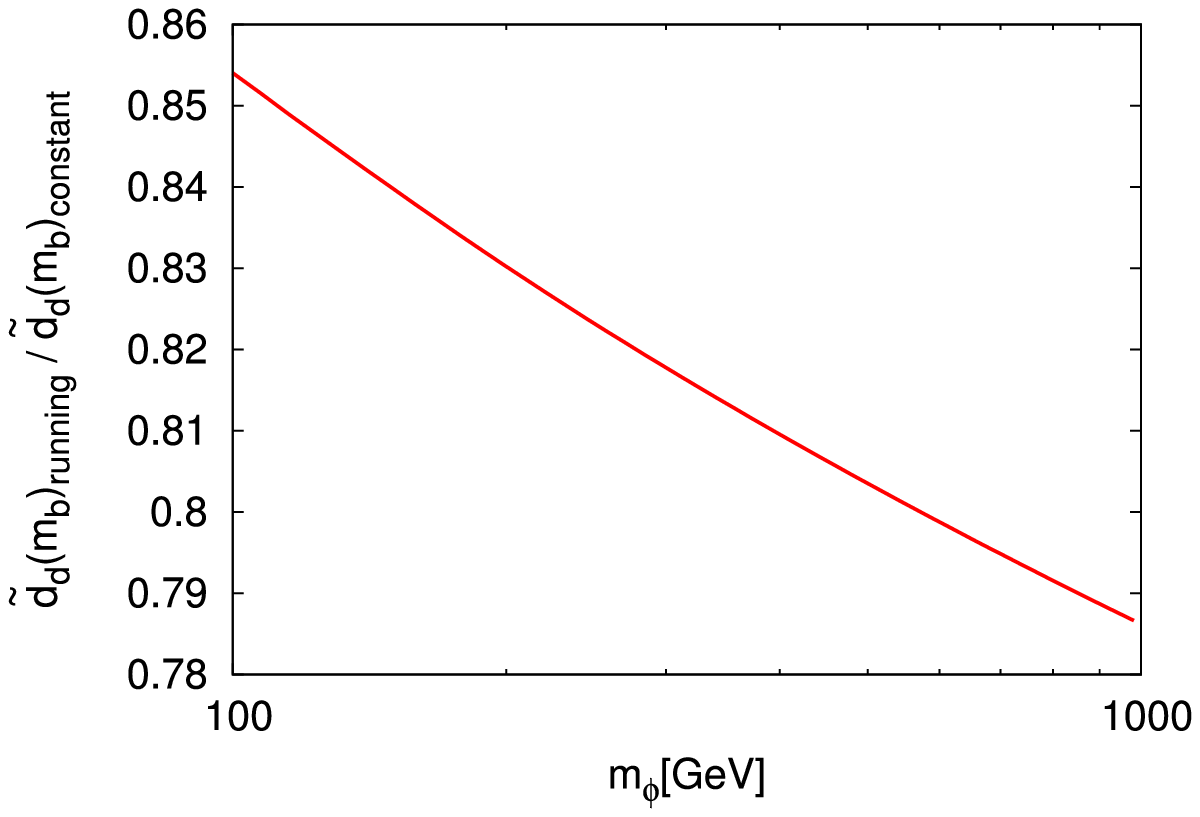} &
   \includegraphics[width=8cm,clip]{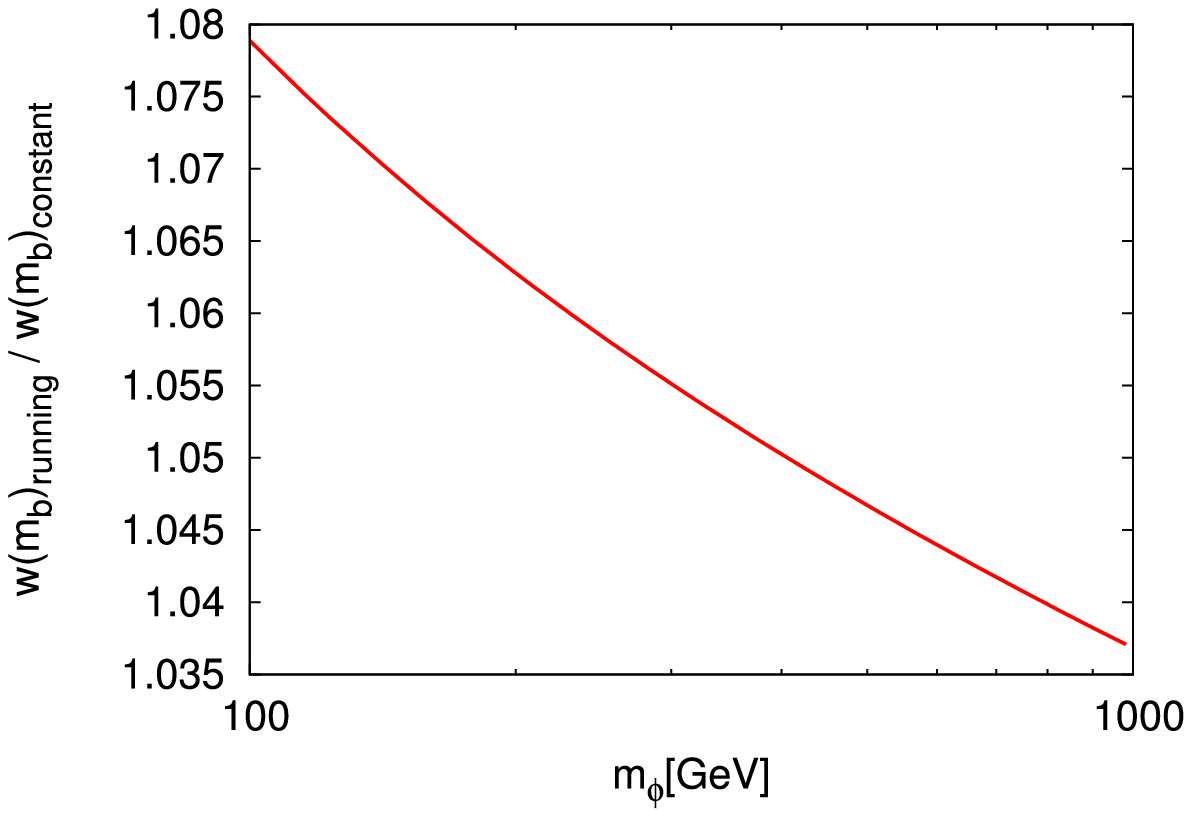} \\
   (a) & (b) 
\end{tabular} 
\caption{(a) : Ratio of the CEDM for down quark, $\tilde{d}_d$, at $\mu=m_b$
  between including and not including running of the strong coupling
  constant, as a function of $m_\phi^{}$. (b) The same ratio for coefficient of  
  three-gluon operator, $w$.}
\end{center}
\end{figure}

\begin{figure}[h]
\begin{center}
   \includegraphics[width=8cm,clip]{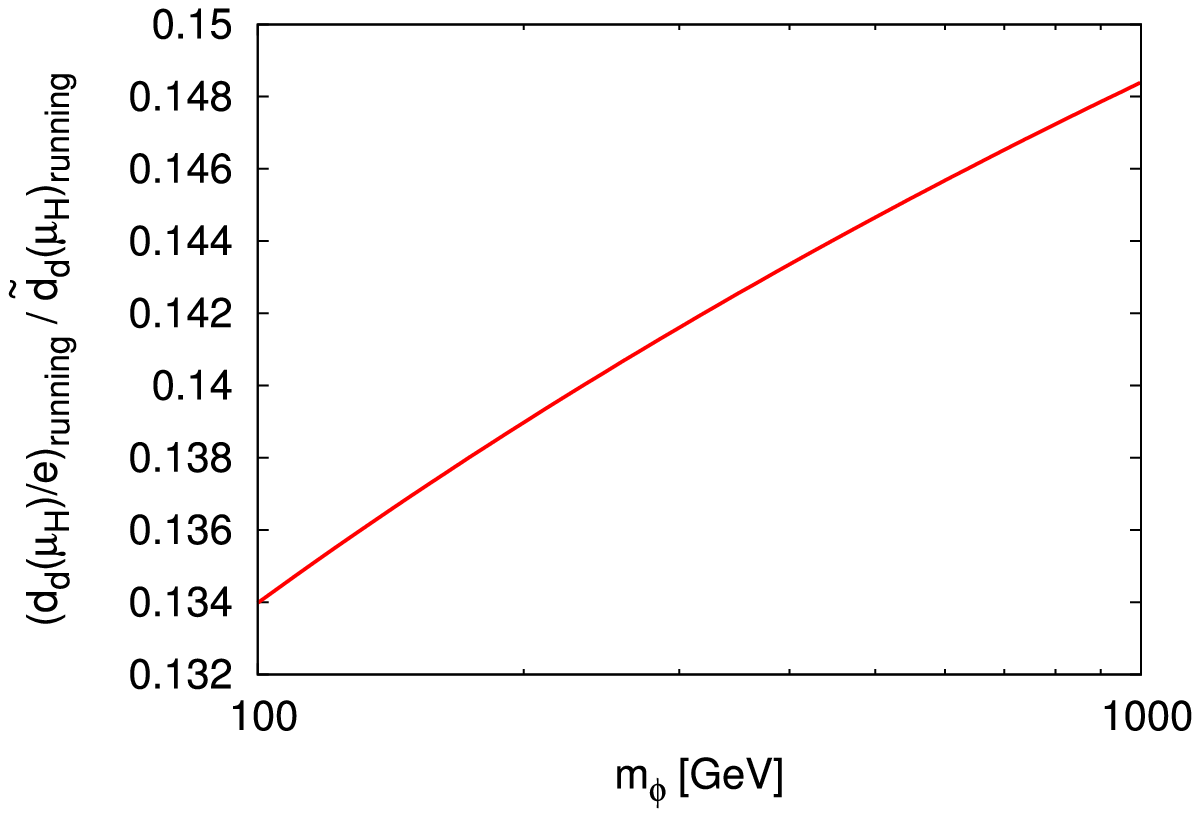} 
\caption{Ratio of EDM and CEDM  for down quark, $(d_d/e)/ \tilde{d}_d$, at hadron scale as a function of $m_\phi^{}$.}
\end{center}
\end{figure}


\section{Color-octet Scalar Boson Exchange}

In the previous section it is shown that the neutral scalar  boson does
not generate sizable EDMs for light quarks via two-loop diagrams at
$O(\alpha_s)$. This comes from cancellation between contributions via
$\widetilde{C}_3^{q'q}$ ($\widetilde{C}_3^{qq'}$) and
$\widetilde{C}_4^{q'q}$ ($\widetilde{C}_4^{qq'}$). This situation is
different when the scalar boson has color. 
Now let us assume that a color-octet
scalar boson $\Sigma(=\Sigma^AT^A)$ is introduced and it has CP-violating Yukawa
interactions with quarks,
\begin{align}
{\mathcal L}_\Sigma 
&= 2^{1/4}G_F^{1/2} m_q
\overline{q_\alpha} (f_S^q + i\,f_P^q \gamma_5) q_\beta \Sigma_{\alpha\beta}\, . 
\label{yukawa2}
\end{align}
The octet scalar fields may appear in new physics beyond the SM such as 
a radiative seesaw model for the neutrino masses~\cite{Ref:OctetNu}, 
and a grand unified model~\cite{Ref:OctetGUT}. 

Integration of the $\Sigma$ leads to the following Wilson coefficients,
\begin{align}
&C_4^q= - \sqrt2G_F 
\frac{m_q^2}{m_{\Sigma}^{2}} (\frac14 +\frac1{2N})f_{S}^{q} f_{P}^{q}  \,, 
&&C_5^q=- \sqrt2G_F 
\frac{m_q^2}{m_{\Sigma}^{2}} \frac1{16}  f_{S}^{q} f_{P}^{q}  \,,\nonumber\\
&C_1^{q'q}= -\sqrt2G_F \frac{ m_qm_{q'}}{m_{\Sigma}^{2}}\frac1{2N} f_{S}^{q'} f_{P}^{q}  \,, 
&&C_2^{q'q}= \sqrt2G_F\frac{ m_qm_{q'} }{m_{\Sigma}^{2}}\frac12 f_{S}^{q'} f_{P}^{q}\,,\nonumber\\
&C_1^{qq'}= -\sqrt2G_F\frac{ m_qm_{q'}}{m_{\Sigma}^{2}} \frac1{2N} f_{S}^{q} f_{P}^{q'}  \,, 
&&C_2^{qq'}= \sqrt2G_F\frac{ m_qm_{q'}}{m_{\Sigma}^{2}}\frac12  f_{S}^{q} f_{P}^{q'} \,.
\end{align}
Now $C_2^{qq'}$ and/or $C_2^{q'q}$ are generated. At the two-loop
level, the EDMs for light quarks have contributions via
$\widetilde{C}_3^{q'q}$ ($\widetilde{C}_3^{qq'}$) and
$\widetilde{C}_4^{q'q}$ ($\widetilde{C}_4^{qq'}$) with different
weights. This is different from the case of the neutral scalar boson exchange
as discussed in the previous section. The EDMs for light quarks are
generated at two-loop level of $O(\alpha_s)$ as
\begin{align}
C_1^q
&= -\frac{\alpha_s}{4\pi^3} \frac{m_{q'}}{m_q} \frac{Q_{q'}}{Q_q} C_{F} \Big(\ln\frac{m_\Sigma}{m_{q'}}\Bigr)^2
\Bigl[\widetilde{C}_2^{q'q}+\widetilde{C}_2^{qq'}\Bigr]\, ,
\end{align}
which is compared with the CEDMs for light quarks as 
\begin{align}
C_2^q
&= \frac{\alpha_s}{8\pi^3} \frac{m_{q'}}{m_q} \Big(\ln\frac{m_\Sigma}{m_{q'}}\Bigr)^2
\Bigl[-\widetilde{C}_1^{q'q} + (\frac{1}{2N}-C_{F}) \widetilde{C}_2^{q'q}
-\widetilde{C}_1^{qq'}
 + (\frac{1}{2N}-C_{F}) \widetilde{C}_2^{qq'}\Bigr]\, .
\end{align}

In Fig.~4 the EDM and CEDM for down quark at the hadron scale are
shown as functions of $m_\Sigma^{}$ with $f_S^d=f_P^d=1$ and
$f_S^b=f_P^b=1$\footnote{ 
If the electrically charged color-octet scalar boson exists, 
we may have large contributions from the top quark~\cite{Ref:OctetGUT}. 
}. Here we use the running coupling for $\alpha_s$. 
It is found that the EDM contribution is larger than the CEDM ones in the
neutron EDM when we adopt the QCD sum rule result on the neutron EDM
evaluation. (See Eq.~(\ref{sumrule}).)

\begin{figure}[h]
\begin{center}
\begin{tabular}{cc}
   \includegraphics[width=8cm,clip]{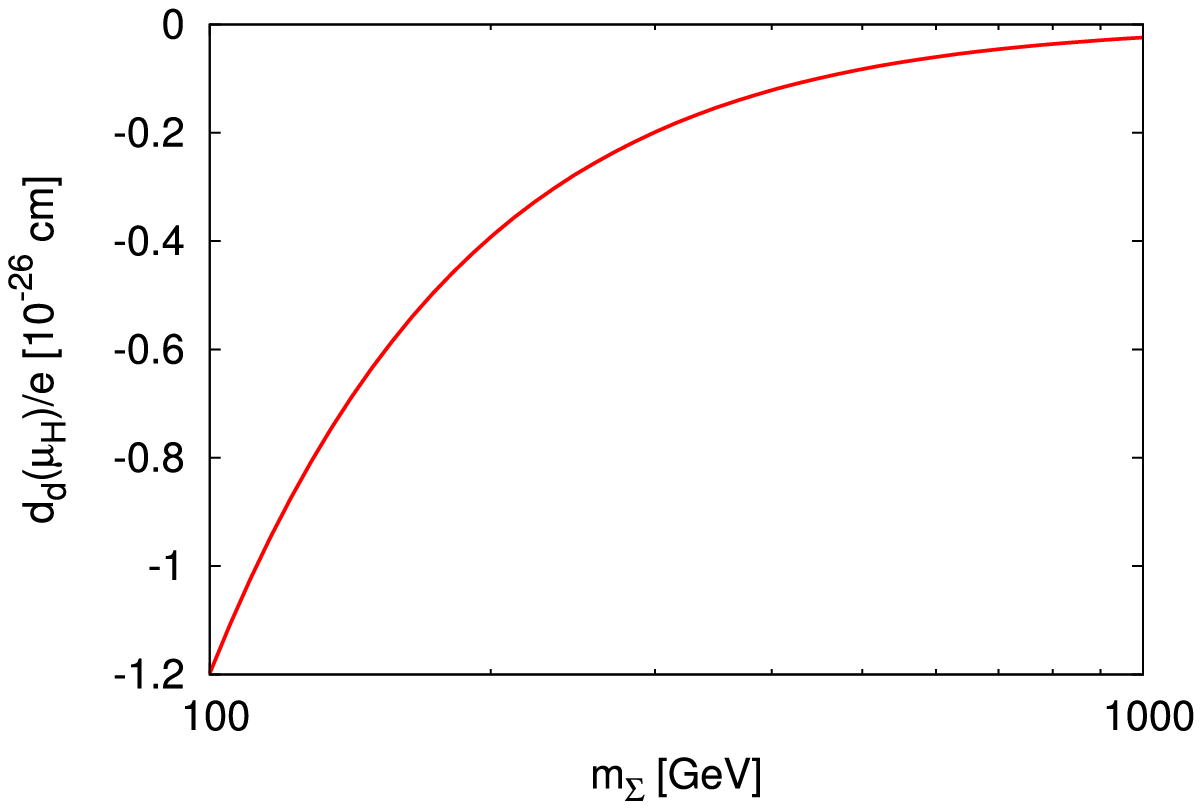} &
   \includegraphics[width=8cm,clip]{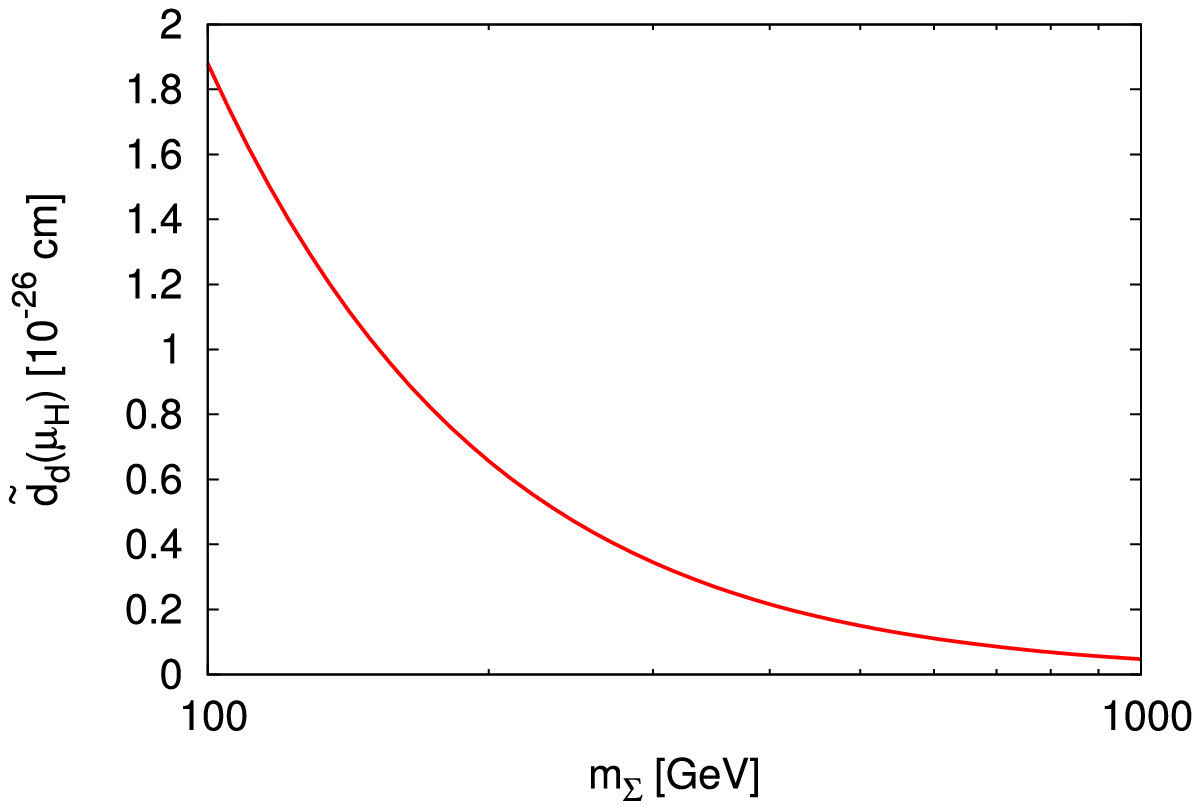} \\
   (a) & (b) 
\end{tabular} 
\caption{(a) EDM $d_d/e$ and (b) CEDM $\tilde{d}_d$ for down quark at hadron scale as 
functions of $m_\Sigma^{}$.}
\end{center}
\end{figure}

For completeness, we also show the magnitude of the three-gluon operator in Fig.~5.
Here, we ignore the short-distance contributions to the CEDMs for heavy
quarks and three-gluon operator whose loop momenta are around
$m_\Sigma$ and take $C_2^q(m_\Sigma^{})=C_3(m_\Sigma^{})=0$ for
simplicity\footnote{
The short-distance contribution to the three-gluon
operator could come from a diagram in which three gluons are emitted from
each quark and scalar boson line, and the evaluation is beyond the scope
of this work. 
}. Again, it is found that the three-gluon operator might give a comparable effect to other contributions. 

\begin{figure}[h]
\begin{center}
  \includegraphics[width=8cm,clip]{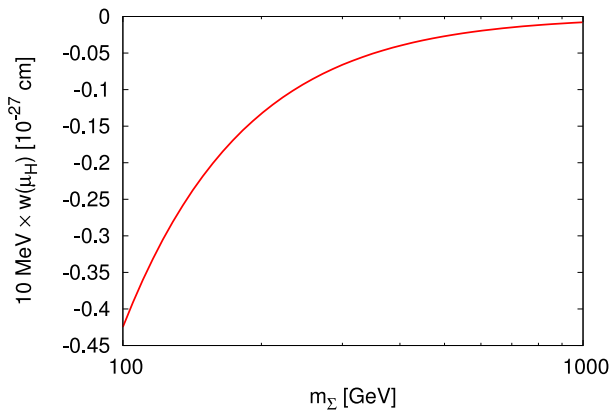} 
  \caption{Three-gluon operator $w$ at hadron scale as a function of
    $m_\Sigma^{}$.}
\end{center}
\end{figure}


\section{Conclusion}

In this Letter, we have derived the renormalization-group equations for
the CP-violating interaction including the quark EDMs and  CEDMs and the
Weinberg's three-gluon operator as well as all the flavor-conserving
four fermion operators.  The operator mixings between the (C)EDM
operators and the four-quark operators are arisen at the order of
$\alpha_s^0$, which can give large contributions to the EDMs via the
renormalization-group evolution. 

Assuming the CP-violating Yukawa interactions for the neutral scalar
bosons, it is known that the CEDMs for light quarks are generated
from the diagrams with heavy-quark loops, called as the Barr-Zee
diagrams. We show that when the neutral scalar boson is much heavier than
heavy quarks, the Barr-Zee diagrams are systematically evaluated
with the RGEs of the CP-violating interaction.  We also show that 
the running effect of the strong coupling constant gives corrections to
the contribution with more than 20 \% compared with assuming the
constant coupling. 
The uncertainties in the calculation of the neutron EDM have been estimated in the literature~\cite{Ref:Uns}. 
It gives about 50 \% error for the QCD sum rule, while 40 \% error for the low-energy constant evaluated from the lattice QCD calculation.
Therefore, hadronic uncertainties would overcome the QCD corrections from the renormalization-group evolution 
at this moment. 
We hope that the lattice QCD simulation will improve and reduce uncertainties significantly~\cite{Ref:Lat}. 

The Barr-Zee diagrams at two-loop level do not contribute to the EDMs
for light quarks at $O(\alpha_s)$. We show using the RGEs of the
CP-violating interaction of a color-singlet scalar boson with quarks
that ratio of the quark EDM over the CEDM is about 0.14, and 
it is roughly proportional to $\log m_\phi^{}$. 
Thus, the contribution is not negligible to the neutron EDM at all.
When the color-octet scalar boson  has CP-violating Yukawa interaction with 
quarks, the quark EDMs are  generated at two-loop level, and they are
comparable to the quark CEDMs.

\section*{Acknowledgments}

This work is supported by Grant-in-Aid for Scientific research from
the Ministry of Education, Science, Sports, and Culture (MEXT), Japan,
No. 20244037, No. 20540252, No. 22244021 and No.23104011 (JH), and
also by World Premier International Research Center Initiative (WPI
Initiative), MEXT, Japan. 
The work of MJSY is supported in part by JSPS Research Fellowships for Young Scientists.




\end{document}